\documentclass[aps,prl,twocolumn,showpacs,amsmath,amssymb,groupedaddress]{revtex4-1}
\usepackage{graphicx}
\usepackage{dcolumn}
\usepackage{bm}
\usepackage{color}
\usepackage{longtable}
\usepackage{multirow}

\begin{document}
\title{The dual nature of magnetism in a uranium heavy fermion system}

\author{Jooseop Lee$^{1,2}$}  \altaffiliation{Formerly with Quantum Condensed Matter Division, ORNL.}
\author{Masaaki Matsuda$^{3}$}
\author{John A. Mydosh$^{4}$}
\author{Igor Zaliznyak$^{5}$}
\author{Alexander I. Kolesnikov$^{3}$}
\author{Stefan S\"ullow$^{6}$}
\author{Jacob P. C. Ruff$^{1}$}
\author{Garrett E. Granroth$^{3}$}

\affiliation{$^1$CHESS, Cornell University, Ithaca, NY 14853, USA}
\affiliation{$^2$CALDES, Institute for Basic Science, Pohang 37673, Korea}
\affiliation{$^3$Neutron Scattering Division, Oak Ridge National Laboratory, Oak Ridge, Tennessee 37831, USA}
\affiliation{$^4$Kamerlingh Onnes
Laboratory and Lorentz Institute, Leiden University, 2300 RA Leiden, The
Netherlands}
\affiliation{$^5$CMPMSD, Brookhaven National Laboratory, Upton, New York 11973, USA}
\affiliation{$^6$Technische Universit\"at Braunschweig, Braunschweig, Germany}

\date{\today}

\begin{abstract}
The duality between localized and itinerant nature of magnetism in $5\textit{f}$ electron systems has been a longstanding puzzle. Here, we report inelastic neutron scattering measurements, which reveal both local and itinerant aspects of magnetism in a single crystalline system of UPt$_{2}$Si$_{2}$. In the antiferromagnetic state, we observe broad continuum of diffuse magnetic scattering with a resonance-like gap of $\approx$ 7 meV, and surprising absence of coherent spin-waves, suggestive of itinerant magnetism. While the gap closes above the Neel temperature, strong dynamic spin correlations persist to high temperature. Nevertheless, the size and temperature dependence of the total magnetic spectral weight can be well described by local moment with $J=4$. Furthermore, polarized neutron measurements reveal that the magnetic fluctuations are mostly transverse, with little or none of the longitudinal component expected for itinerant moments. These results suggest that a dual description of local and itinerant magnetism is required to understand UPt$_{2}$Si$_{2}$, and by extension, other 5$f$ systems in general.
\end{abstract}


\maketitle


The degree of localization of magnetic moments is an important concept for understanding many exotic phenomena in condensed matter, thereby creating the ``duality'' problem \cite{Moriya1985}. The situation is even more complex in multi-band electronic systems, where the localization can be orbital-selective. For example, the magnetism in iron-based superconductors has been long discussed in terms of either itinerant or local moment only models. Recent progress in this field, however, suggests that this system belong to the intermediate coupling region with U/W $\approx$ 1 (U: Coulomb repulsion, W: bandwidth) where we do not have a good understanding yet even for a single band \cite{Dai}.

Ternary intermetallic uranium compounds UT$_{2}$M$_{2}$ (T: a transition metal, M: Si or Ge) have been of great interest in strongly correlated electron physics during the last decades. URu$_{2}$Si$_{2}$ has a very small magnetic moment and shows the famous, yet-to-be-understood, phenomena of the hidden order and unconventional superconductivity \cite{Mydosh}. UPt$_{2}$Si$_{2}$, on the other hand, has been long considered a rare example of uranium intermetallic compound with strongly localized $\textit{f}$-electrons. It orders antiferromagnetically (T$_{N}$=32 K) along c-axis with a magnetic moment of $\approx$ 2 $\mu_{B}$/U \cite{Steeman1990}. Early studies suggested that magnetic anisotropy, high field magnetization, as well as temperature dependence of magnetic susceptibility can be well described within a local-moment crystal electric field (CEF) model \cite{Nieuwenhuys, Yamagishi}.

Recent high field studies \cite{Grachtrup2012, Grachtrup2017}, however, question the degree of localization in this system, suggesting that the observed phase transitions under applied magnetic field can be understood as Lifshitz transitions, an abrupt change in the topology of a Fermi surface. This view is further supported by the density functional theory (DFT) \cite{Elgazzar}, which indicates that 5\textit{f} electrons in UPt$_{2}$Si$_{2}$ system are orbitally polarized and mostly itinerant, with only a slight tendency toward localization.

In order to understand the magnetism, specifically the interplay between local and itinerant nature of the moments in this system, it is crucial to study the spin dynamics. However, magnetic excitations in UPt$_{2}$Si$_{2}$ have not been observed, despite the large ordered magnetic moment. An early inelastic neutron scattering study \cite{Steeman1988} provided very limited information (at 77 K) due to the polycrystalline nature of the sample, while a more recent work \cite{Metoki} did not find any spin waves below $\sim 3$ meV.

\begin{figure*}[tp]
\centering
\includegraphics[width=1.0\hsize]{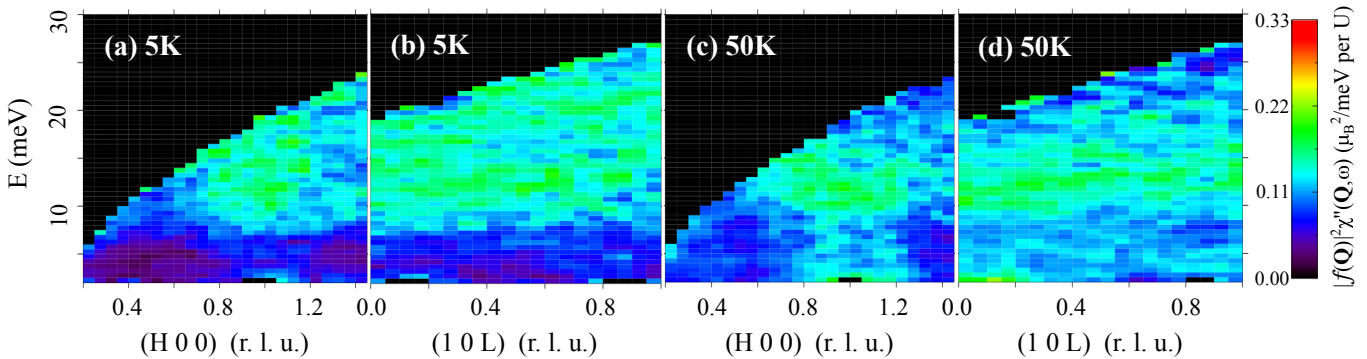}
\caption{The imaginary part of the dynamical magnetic susceptibility, $\chi''(\textbf{Q},\omega)$, corrected for U$^{4+}$ magnetic form factor, at 5 K ((a) and (b)) and 50 K ((c) and (d)). The measurement was performed at SEQUOIA, with incident energy $\textrm{E}_i = 50$ meV.}
\label{fig:SEQdata}
\end{figure*}

Here, we present comprehensive neutron inelastic scattering results demonstrating that both itinerant and local moments nature of \textit{f}-electrons are playing a role in this system. We observe a diffuse magnetic excitation continuum with a resonance-like gap of $\approx$ 7 meV, which clearly cannot be explained by spin wave theory for coherent collective excitation of localized moments. Rather, the excitation can be understood within the random-phase approximation (RPA) model response of the itinerant system. The size and temperature dependence of the total magnetic moment, however, can be well described by local model with $J=4$, a total angular momentum associated with a local magnetic moment, $\hat{\mbox{\boldmath$\mu$}} = g \mu_B \hat{\mbox{\boldmath$J$}}$. Polarized neutron measurements also reveal that the fluctuations are mostly transverse to the staggered ordered moment. While an ordered local moment produces a transverse fluctuation, an itinerant moment also has longitudinal dynamics corresponding to the fluctuation of the size of the magnetic moment \cite{Moriya1985}. These observations show that even in this large magnetic moment system a dual approach based on itinerant and local description is necessary to fully capture its nature. The duality of magnetism was also suggested in other heavy fermion superconductors \cite{Fujimori}, and thus seems to be universal across $5\textit{f}$ electron systems.


All measurements presented here were performed on a 1.5 g single crystalline sample of UPt$_{2}$Si$_{2}$ \cite{Supple1}. Initial neutron scattering measurements were done with the HB1 thermal triple axis spectrometer (TAS) at the HFIR, ORNL. A large volume of energy-momentum space was then explored using the time-of-flight (TOF) spectrometer SEQUOIA \cite{Granroth2006,Granroth2010} at SNS, ORNL \cite{mason2006spallation} with incident energies ($\textrm{E}_{i}$) of 50 meV and 100 meV. TOF neutron data was normalized to the absolute scattering cross-section in $\mu_{B}^{2}/\textrm{meV}/\textrm{U}$ by using a standard vanadium sample \cite{Xu}. The measured scattering intensities were converted to the imaginary part of the dynamical magnetic susceptibility via the fluctuation-dissipation theorem, $\chi''\left(\textrm{\textbf{Q}},\omega\right)= \pi (1-e^{-E/k_{B}T}) S(\textbf{Q},\omega)$ \cite{ZaliznyakLee}.



Contrary to what is expected for conventional magnetic order, a broad continuum of magnetic excitations is observed around the magnetic wave vector, $\textrm{\textbf{Q}}_{M}$=(1 0 0), Fig. \ref{fig:SEQdata}. In the local magnetism approach, interactions between local moments are described by a spin exchange Hamiltonian.  The low-energy collective transverse fluctuations of ordered magnetic moments, i.e., spin waves are expected to have well-defined sharp dispersion relations demonstrating long-range coherence.  In an itinerant moment system, on the other hand, the excitations originate from electron-hole pairs created across the Fermi surface.  These excitations are usually broad in Q-E space and weak in intensity, without clear dispersion.

Figure \ref{fig:SEQdata} (a,b) presents the generalized susceptibility as a function of wave vector along (H 0 0) and (1 0 L) direction, respectively, at T$ \ll$~T$_{N}$. The excitation is diffuse, centered around E $\approx$ 13 meV, with a gap $\approx$ 7 meV. The intensity in H-direction is peaked near $\textrm{\textbf{Q}}_{M}$. The excitation along L-direction (Fig. \ref{fig:SEQdata} (b)), in contrast, is rather flat, suggesting anisotropic magnetic interactions consistent with the quasi two-dimensional (2D) character observed in resistivity \cite{Sullow} and Fermi surface \cite{Elgazzar}. A spin gap of $\sim$7 meV, which is clearly visible in Fig.~\ref{fig:SEQdata} (a,b) is roughly consistent with the gap of 46 K estimated from the temperature dependence of resistivity \cite{Sullow}. Above the Neel temperature, at 50 K, this gap is closed, as shown in Fig.~\ref{fig:SEQdata} (c,d). It should be noted that spin fluctuations along (H 0 0) are still clearly peaked around $\textrm{\textbf{Q}}_{M}$, indicating strong magnetic correlation even in the paramagnetic regime above T$_{N}$. We refer to the supplementary material for more temperature dependence data \cite{Supple6}.

As the observed excitation continuum cannot be described by spin wave theory, we fit the data using the dynamical magnetic susceptibility calculated using Random Phase Approximation (RPA) in an itinerant electron model. The resulting expression is essentially identical to the extended Self-Consistent Renormalization (SCR) theory model \cite{Moriya1974, Moriya1987, Moriya1985}, which takes account, in a self-consistent way, of the effect of the spin fluctuation mode coupling\cite{Bao1998}. The resulting RPA-SCR expression for the generalized magnetic susceptibility is,

\begin{equation}
\chi^{\prime \prime}( \textbf{Q+q}, \omega ) = \frac{\chi_{Q}}{1 + (q/\kappa)^{2}}  \frac{\hbar\omega\Gamma (q,\kappa)}{(\hbar\omega)^{2}+\Gamma(q,\kappa)^{2}} \, ,
\label{eq:SCRtheory}
\end{equation}
with the Q-dependent relaxation rate,

\begin{equation}
\Gamma(q,\kappa) = \gamma_{A}\left( \kappa^{2} + q^{2} \right)
\label{eq:relaxation}.
\end{equation}

Parameters $\kappa$, $\gamma_{A}$, and $\chi_{Q}$ are a characteristic width in reciprocal space, a temperature-insensitive energy width parameter, and a static staggered susceptibility, respectively. The wave vector $q$ is measured away from $\textrm{\textbf{Q}}_{M}$.

\begin{figure}[tp]
\centering
\includegraphics[width=1.0\hsize]{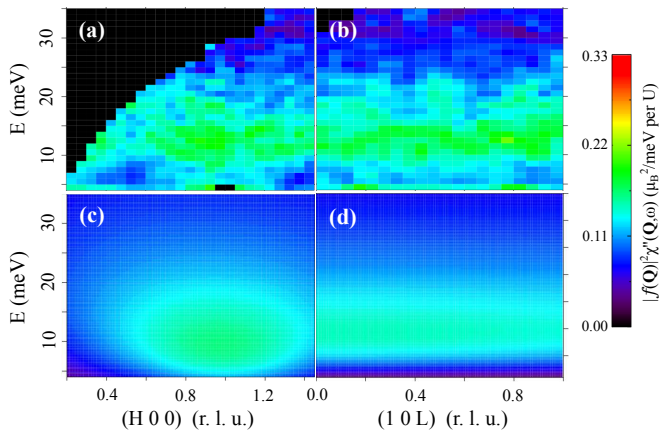}
\caption{(a) and (b)  $\chi''\left(\textrm{\textbf{Q}},\omega\right)$ observed at 50K with $\textrm{E}_i=100$ meV along H and L direction, respectively. (c) and (d) Fit to SCR, or equivalently, RPA theoretical model.}
\label{fig:SCRfit}
\end{figure}

It is clear from Fig. \ref{fig:SCRfit} that Eq. \ref{eq:SCRtheory} provides adequate description of the observed magnetic excitation. While data with $\textrm{E}_i=50$ meV reveals the 7 meV gap with better resolution, $\textrm{E}_i=100$ meV data capture the full range of magnetic excitation and, therefore, was used for the global fitting. The results of the fit along the H and L directions are shown in Fig. \ref{fig:SCRfit} (c,d). Table \ref{tab:SCRparm} lists the best fit parameters when magnetic form factor is co-refined \cite{Comment0, Supple2}. The top of the excitation band is around 25 meV. The single intensity lobe is inconsistent with dispersive magnetic excitations. This demonstrates that the spin fluctuation is rather a resonance localized in Q-E space, similar to the resonance magnetic excitation observed in many unconventional superconductors such as cuprates \cite{Cuprate}, Fe-based superconductors \cite{Dai}, or heavy fermion superconductors \cite{Petrovic}

\begin{table}[bp] 
\centering
\begin{tabular}{c c c c }
\hline\hline
& ~$\kappa$~$[\textrm{\AA}^{-1}]$~ & ~$\gamma_{A}$~$[\textrm{meV}/\textrm{\AA}^{-2}]$~ & ~$\chi_{Q}$~$[\mu_{B}^{2}/\textrm{meV}]$~  \\	
\hline
(H 0 0) & 1.35(8) & 5.18(54) & 0.33(1)  \\
(1 0 L) & 2.77(114) & 1.21(98) & 0.30(1)  \\
\hline\hline
\end{tabular}
\caption{Parameters obtained by fitting $\textrm{E}_i$=100 meV and T=50 K data with Eq. \ref{eq:SCRtheory}.}
\label{tab:SCRparm}
\end{table}

\begin{figure}[tbp]
\centering
\includegraphics[width=0.7\hsize]{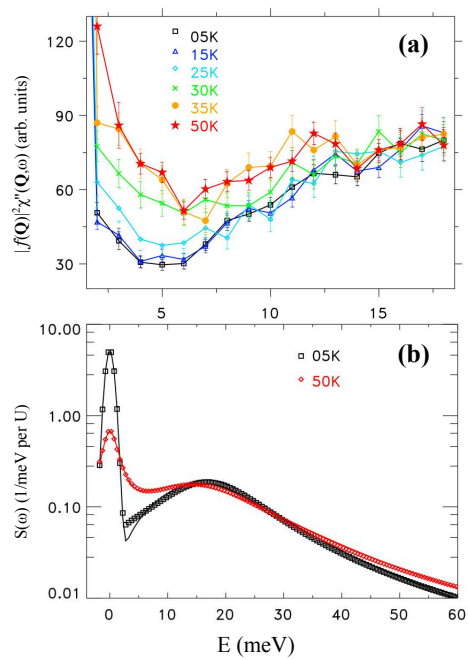}
\caption{Evolution of the generalized susceptibility across T$_N$. (a) The detailed temperature dependence of low energy spin excitation at Q=(1 0 0) measured using HB1 triple axis spectrometer. Counting time for each data point was about 5 minutes. (b) The autocorrelation function, S(E), at 5 K (black square) and 50K (red diamond), obtained from Q-integrated spectral weight of the TOF data. Solid lines show fit to a model function consisting of a Lorentzian centered at E = 0 (quasi-elastic) and a damped harmonic oscillator (inelastic). }
\label{fig:TASdata}
\end{figure}

Figure \ref{fig:TASdata} shows the change in scattering intensity with varying temperature.  There is an increase of quasi-elastic paramagnetic scattering with the increasing temperature across T$_{N}$. The change of the spectral gap follows the order parameter dependence of magnetic Bragg peaks \cite{Steeman1990}, which strongly supports the magnetic nature of fluctuations \cite{Sergeicheva}.

The spectral weight filling in the gap comes from other energy transfers, Fig. \ref{fig:TASdata} (b). It is clear, however, from Figs. \ref{fig:SEQdata} and \ref{fig:TASdata}, that this redistribution affects only a moderately small fraction of the total magnetic spectral weight. In this case, the ordered magnetic moment in the antiferromagnetic state, which is only 2 $\mu_{B}$, is weak and comprises only a small fraction of the fluctuating magnetic moment. Consequently, its influence on magnetic excitations is small and limited to low energies,  $ \lesssim k_{\textrm{B}}\textrm{T}_{\textrm{N}}$, which again emphasizes that the system cannot be simply understood from its ordered moment \cite{Supple3}. 

In the local moment model, the integral spectral weight measured by neutron scattering obeys the zero-moment sum rule \cite{ZaliznyakLee}, $\sum_\alpha \int_{-\infty}^{+\infty}\int_{BZ} S^{\alpha\alpha}(\textbf{q},\omega) d\textbf{q}dE ~/ \int_{BZ} d\textbf{q} = \left(g \mu_B \right)^2 J(J+1)$, where $\alpha$ is the polarization of magnetic fluctuation, $g$ is the spectroscopic Lande g-factor, $\mu_B$ is Bohr magneton. By assuming random polarization of magnetic fluctuation with respect to the wave vector transfer, which is a good approximation for the TOF data in Fig. \ref{fig:TASdata} (b), and fitting the Q-integrated normalized magnetic intensity \cite{Comment1}, $S(E)$, to a model function consisting of the quasielastic Lorentzian and the damped harmonic oscillator (DHO), we obtain the integral magnetic scattering intensity of $\approx 15.9 \mu_{B}^{2}$ at 5 K and $\approx 13.6 \mu_{B}^{2}$ at 50 K. Using $g=0.8$, which is appropriate for the $^{3}H_{4}$ Hund rule Russell-Saunders ground state of U$^{4+}$ \cite{MFF}, this results in estimated $J\approx 4.5$ at 5 K and $\approx 4.1$ at 50 K, consistent with the $J=4$ state. This exhausts the full magnetic spectral weight available for the 5$f^2$ electronic configuration of U$^{4+}$ and, therefore, temperature enhancement of integral spectral weight from the entanglement between local and itinerant electrons \cite{IgorPRL} is not observed. 

The integral intensity of the magnetic excitation spectrum thus indicates participation of two $5f$ electrons, as in the $J=4$ state of U$^{4+}$ in the local-moment picture. In order to rationalize the observed magnetic spectral weight in the itinerant-electron model of Eq. \ref{eq:SCRtheory}, one needs to recall that the approximation adopted in deriving this result \cite{Moriya1974,Moriya1987,Moriya1985} limits its applicability to the proximity of the Fermi energy. Hence, the integration of the spectral weight must be limited by a finite-energy cut-off, $\Lambda$, which is of the order of the itinerant-electron bandwidth (otherwise the integral formally diverges). On account of this cutoff, the total magnetic spectral weight is, $\int_{-\infty}^{+\infty}\int_{BZ} \chi'' (\textbf{q},\omega) d\textbf{q}dE ~/ \int_{BZ} d\textbf{q} \approx \chi_Q (\gamma_A \kappa^2) \ln \left( \frac{\Lambda}{\gamma_A \kappa^2} \right)$ \cite{Supple4}. Using the fit results from Table \ref{tab:SCRparm} we estimate the itinerant-electron bandwidth of 1 $\sim$ 2 eV, in good agreement with the extent of spin-polarized 5$f$ bands in recent DFT calculations \cite{Elgazzar,Grachtrup2017}. Hence, UPt$_2$Si$_2$ cannot be simply viewed as narrow-band system where local-moment picture applies. The observed magnetic dynamics inside the energy window probed in the present experiment is distinct from that of local moments and is well described by the RPA itinerant-electron theory. On the other hand, applying the first moment sum rule to the measured intensity and assuming local moment model with nearest neighbor exchange interactions we obtain a much lower energy scale, $\approx$ 10 meV, for the contribution of magnetic bond energies per U to the ground state, consistent with the observed excitation spectrum \cite{Supple5}.

\begin{table}[bp] 
\centering
\begin{tabular}{c c c}
\hline\hline
& ~~~HF (\textbf{P$_{n}$} $\mathbin{\!/\mkern-5mu/\!}$ a)~~~ & ~~~VF (\textbf{P$_{n}$} $\mathbin{\!/\mkern-5mu/\!}$ b)~~~\\	
\hline
SF  & S$_{b}^{2}$ + S$_{c}^{2}$ + B$_{1}$ & S$_{c}^{2}$ + B$_{1}$ \\
NSF & N$^{2}$ + B$_{2}$ & S$_{b}^{2}$ + N$^{2}$ + B$_{2}$ \\
\hline\hline
\end{tabular}
\caption{Nuclear and magnetic components contributing to the scattering intensity at (H 0 0) in the present polarized neutron measurement. $P_{n}$ is the neutron spin polarization direction, $N$ denotes nuclear components, and B$_{1, 2}$ represent background in SF and NSF configuration, respectively. $S_{a}$ cannot be observed in this setup because magnetic component parallel to $\textbf{Q}$ does not contribute to neutron scattering.}
\label{tab:PolarSetup}
\end{table}

\begin{figure}[tp]
\centering
\includegraphics[width=0.7\hsize]{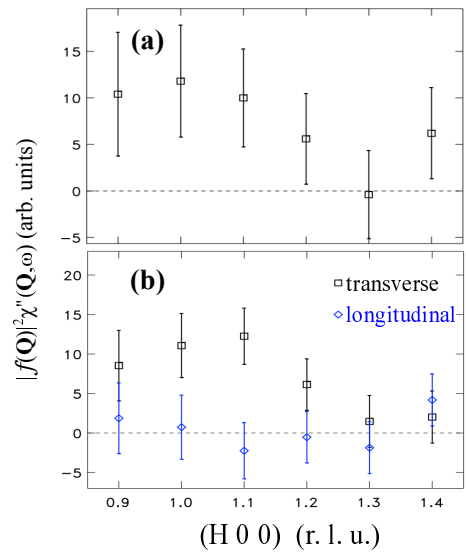}
\caption{Polarized neutron measurement at constant E=12 meV at T=5 K. (a) Magnetic scattering (= HF-SF $-$ HF-NSF in Table \ref{tab:PolarSetup}). (b) Transverse spin fluctuation, $S_{b}$ (=HF-SF $-$ VF-SF or VF-NSF $-$ HF-NSF), and estimated longitudinal spin fluctuation, $S_{c}$ (=HF-SF $-$ VF-NSF or VF-SF $-$ HF-NSF). Each data point has been counted for about 3 hours.}
\label{fig:PTASdata}
\end{figure}

In order to further elucidate the magnetic nature of the observed diffuse and weak excitation, we carried out a polarized neutron measurement. Table \ref{tab:PolarSetup} summarizes different contributions to the spin-flip (SF) and non-spin-flip (NSF) intensities when the scattering plane is (H 0 L) and the neutron spin is parallel to either $a$-axis (HF: horizontal field) or $b$-axis (VF: vertical field) directions. Since U magnetic moments are aligned along $c$-axis, $S_{a, b}$ corresponds to transverse fluctuations while $S_{c}$ indicates a longitudinal fluctuation.

Figure \ref{fig:PTASdata} presents polarized neutron scans below T$_N$ at constant E=12 meV along (H 0 0) direction, which reveal small but clear magnetic signal whose Q-dependence is consistent with that observed with unpolarized neutrons (Fig. \ref{fig:SCRfit}). The polarization of magnetic fluctuation can be further analyzed by comparing different scattering setups. The intensity difference between HF and VF configuration, either in SF or NSF, directly yields $S_{b}$, the transverse fluctuation. Black squares in Fig. \ref{fig:PTASdata} (b) show the transverse fluctuation. The longitudinal component can be evaluated from the difference of HF-SF and VF-NSF (or VF-SF and HF-NSF) assuming that $\textrm{N}^{2}+\textrm{B}_{2}$ and $\textrm{B}_{1}$ do not differ, which is valid since the total magnetic signal (Fig. \ref{fig:PTASdata} (a)) correspond to the sum of transverse and longitudinal signals (Fig. \ref{fig:PTASdata} (b)). The estimated longitudinal fluctuation shown as blue diamond symbols in Fig. \ref{fig:PTASdata} (b) appears to have negligible intensity. This indicates that spin fluctuations are primarily transverse, in sharp contrast with the longitudinal magnetic dynamics observed in chromium, the archetypal example of an itinerant antiferromagnet \cite{Chromium} and in other uranium compounds such as UN \cite{UN} and URu$_2$Si$_2$ \cite{Flouquet}.\\


Our comprehensive inelastic neutron scattering studies thus show that itinerant electrons are playing a major role in the dynamical magnetism of UPt$_2$Si$_2$. Despite the large ordered magnetic moment revealed by neutron diffraction \cite{Steeman1990}, the magnetic excitations are broad and resonance-like, lacking a sharp dispersion, and could be understood in itinerant RPA approach. Above T$_N$, the gap in the excitation spectrum is closed, but the strong magnetic correlation is still present. The temperature dependence of spectral weight shows that only small fraction of magnetic excitation changes across the Neel temperature and the system should not be described solely from its ordered moment. \\

Our results make it clear that this system is at the boundary where both local moment and itinerant degrees of freedom are important, just like Fe-based superconductors or other strongly correlated electron systems. Many physical properties previously attributed exclusively to local nature, for example, field induced phase transitions \cite{Grachtrup2012} could be indeed consequences of this magnetic duality. Lack of adequate theory to comprehensively describe the observed behaviors call for new modeling effort with UPt$_2$Si$_2$ as a test case for duality. Considering the universal magnetic duality in 5$\textit{f}$-electron systems, other overlooked large moment uranium intermetallics \cite{Rh122,Pd122,Ni122,Cu122} should be also re-examined.


\section{Acknowledment}

We thank Guangyong Xu and T. J. Williams for helpful discussion on data analysis. This research at High Flux Isotope Reactor and Spallation Neutron Source of ORNL was sponsored by the Scientific User Facilities Division, Office of Basic Energy Sciences, U.S. Department of Energy and the Laboratory Directrors Research and Development fund of ORNL. Cornell High Energy Synchrotron Source was supported by the NSF and NIH/National Institute of General Medical Sciences via NSF award no. DMR-1332208. The work at Brookhaven National Laboratory was supported by the Office of Basic Energy Sciences, U.S. Department of Energy, under Contract No. DE-SC0012704.



\begin{thebibliography}{99}
\bibitem{Comment0}
The error bar does not include systemic errors from absolute normalization, which is typically about 20 $\%$.
\bibitem{Comment1}
Since the accessible energy range around Q=(1 0 0) is limited due to kinematic restriction, we divided BZ into small areas whose integrated intensity is extrapolated to high energy transfer by fitting to Lorentzian distribution. Final integrated intensity in Fig. 3(b) was estimated by summing each integrated intensity weighted by their BZ volume. E$_i$=50meV and 100meV data have been used for low ($\leq$ 10meV) and high energy ($\geq$ 10meV) transfer, respectively, which were joined smoothly after absolute normalization.
\end{thebibliography}
\end{document}


\title{Supplementary material for ``the dual nature of magnetism in a uranium heavy fermion system''}

\author{Jooseop Lee$^{1,2}$}  \altaffiliation{Formerly with Quantum Condensed Matter Division, ORNL.}
\author{Masaaki Matsuda$^{3}$}
\author{John A. Mydosh$^{4}$}
\author{Igor Zaliznyak$^{5}$}
\author{Alexander I. Kolesnikov$^{3}$}
\author{Stefan S\"ullow$^{6}$}
\author{Jacob P. C. Ruff$^{1}$}
\author{Garrett E. Granroth$^{3}$}

\affiliation{$^1$CHESS, Cornell University, Ithaca, NY 14853, USA}
\affiliation{$^2$CALDES, Institute for Basic Science, Pohang 37673, Korea}
\affiliation{$^3$Neutron Scattering Division, Oak Ridge National Laboratory, Oak Ridge, Tennessee 37831, USA}
\affiliation{$^4$Kamerlingh Onnes
Laboratory and Lorentz Institute, Leiden University, 2300 RA Leiden, The
Netherlands}
\affiliation{$^5$CMPMSD, Brookhaven National Laboratory, Upton, New York 11973, USA}
\affiliation{$^6$Technische Universit\"at Braunschweig, Braunschweig, Germany}

\maketitle

\section{I. Experimental Setup}

The single crystal UPt$_{2}$Si$_{2}$ was prepared using a modified Czochralski method at Leiden University. The mosaic of the sample is 0.482(4)$^{\circ}$ at Q = (2 0 0) and 0.660(24)$^{\circ}$ at Q = (0 0 4).\\

For the unpolarized and polarized triple axis spectrometer measurement, the beam collimation was 48$'$-80$'$-80$'$-open with Soller collimators around sample and final energy was fixed to E$_f$ = 13.5 meV. The resulting energy resolution is about 1.3 meV at the elastic position. For the polarization measurement, Heusler monochrometer and analyzer were used with a guide field to guide the neutron spin to the horizontal field (HF) and vertical field (VF) directions at the sample position. The flipping ratio measured at Q = (1 0 0) is 15.3(19) in the polarized mode.\\

When using Time-Of-Flight (TOF) neutron spectrometer to cover large energy-momentum space, incident energies of 50 meV and 100 meV were selected by rotating Fermi chopper 1 at a speed of 360 Hz and 480 Hz, respectively. The corresponding energy resolution values at the elastic position are 1.8 meV and 4.4 meV. The crystal was aligned in the [H 0 L] plane and rotated in 1$^{\circ}$ steps over 90$^{\circ}$ range and counted about 10 minutes at each angle. The obtained data was reduced to $\textrm{S}(\mathbf{Q},\omega)$ using Mantid \cite{arnold2014mantid} and data slicing was performed with Mantid and Horace \cite{EWINGS2016132}.\\

The sample was loaded into a aluminum can and mounted to a closed cycle refrigerator at both instruments.\\

\section{II. Magnetic Form Factor}

Magnetic Form Factor (MFF) of U$^{+3}$ and U$^{+4}$ is reported to have negligible difference \cite{MFF} and  it could be reasonable to use U$^{+4}$ as is commonly done for URu$_2$Si$_2$, which is further supported by the experimentally observed total magnetic moment corresponding to $J$ = 4.\\

Here, however, we find the shift of intensity peak in Q-space towards lower-Q when we simulated the magnetic excitation with MFF of U$^{+4}$.  In the absence of precise knowledge of MFF, it is then possible to account for the orbital hybridization effects by using a phenomenological MFF, as it is customary in magnetic neutron diffraction structural refinements and other situations \cite{Igor}. Simply by introducing a scale factor, p, to the momentum transfer of the magnetic form factor obtained in dipole approximation, we can adjust the spatial extent of the electronic magnetization density: $F(Q) = F_{dipole} (pQ)$.\\

We find the scale factor, p $\approx$ 0.5, which suggests that electrons are about half as localized in real space compared with free U$^{+4}$ ions. By fitting magnetic excitation along H-direction, we were able to find p values of 0.46(10) at 5 K and 0.41(10) at 50 K. The fit along L-direction becomes unstable when $\kappa$ (characteristic width in reciprocal space) and MFF are refined together because these two parameters are coupled when correlation length is short and the width is large. In this case, the broad peak can be ascribed to either very short correlation length, or entirely to the magnetic form factor. Here, we fixed the p value to the value obtained in H-direction for fitting the width in L-direction.\\

\section{III. SCR-RPA Analysis in Antiferromagnetically Ordered Phase}

\begin{figure}[htp]
\centering
\includegraphics[width=0.65\hsize]{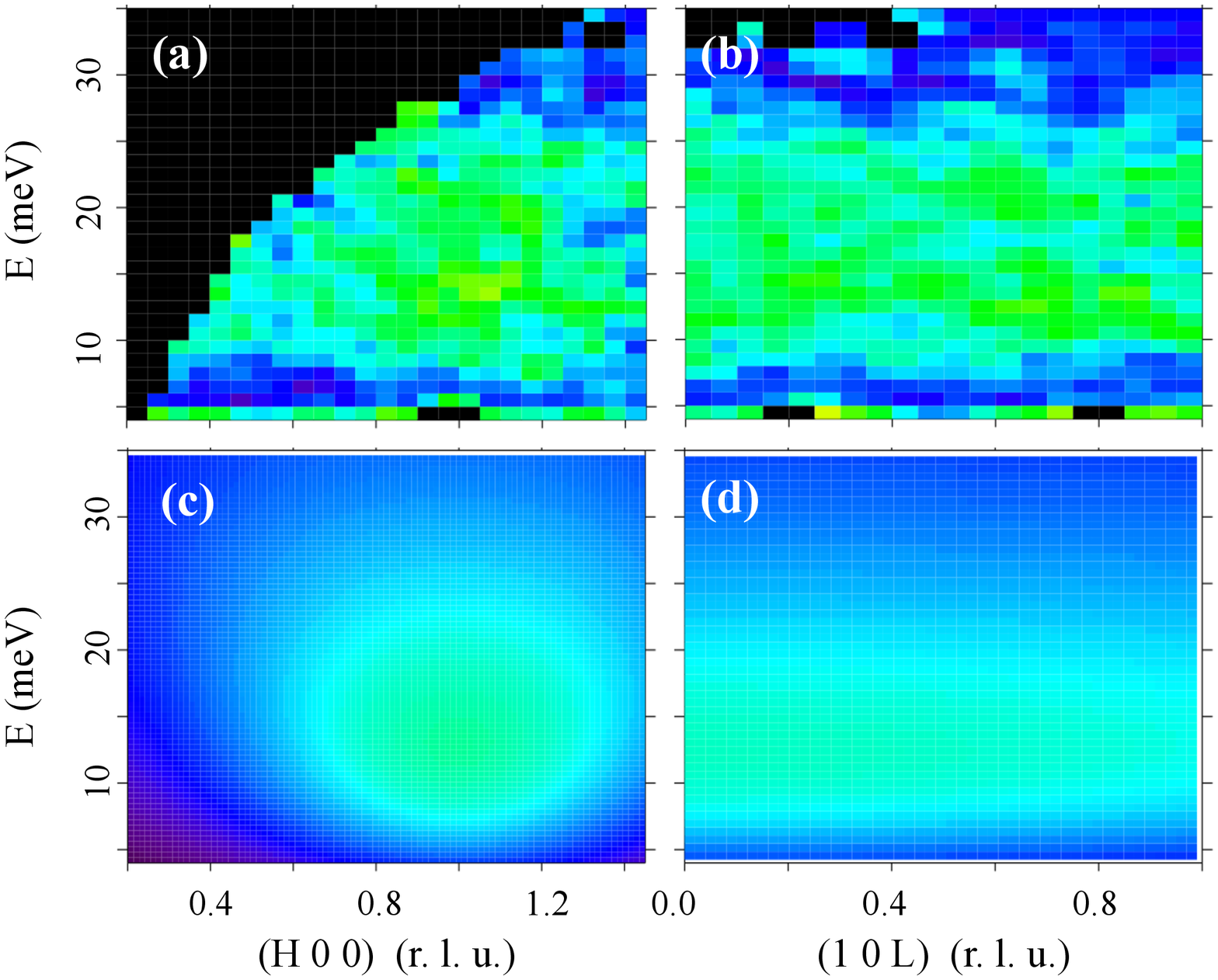}
\caption{(a) and (b)  $\chi''\left(\textrm{\textbf{Q}},\omega\right)$ observed at 5K with $\textrm{E}_i=100$ meV along H and L direction, respectively. (c) and (d) Fit to SCR, or equivalently, RPA theoretical model.}
\label{fig:S1}
\end{figure}

\begin{table}[hbp] 
\centering
\begin{tabular}{c c c c }
\hline\hline
& ~$\kappa$~$[\textrm{\AA}^{-1}]$~ & ~$\gamma_{A}$~$[\textrm{meV}/\textrm{\AA}^{-2}]$~ & ~$\chi_{Q}$~$[\mu_{B}^{2}/\textrm{meV}]$~  \\	
\hline
(H 0 0) & 2.03(15) & 2.28(32) & 0.33(1)  \\
(1 0 L) & 1.21(28) & 4.54(190) & 0.36(2)  \\
\hline\hline
\end{tabular}
\caption{Parameters obtained by fitting $\textrm{E}_i$=100 meV and T=5 K data.}
\label{tab:SCRparm5K}
\end{table}

Table \ref{tab:SCRparm5K} shows fit parameters to RPA-SCR theory in the AFM ordered state at 5 K. Fig. \ref{fig:S1} shows measured data at 5K with $\textrm{E}_i$=100 meV and simulated data from the fit values. Here, we show that there is no drastic change between AFM and PM phases. The difference in fit parameters are mainly from large error coming from weak intensity and featureless-shaped spectrum in Q-E space. As presented in Fig. 1 of main text, comparison of excitation between antiferromagnetic and paramagnetic phases reveals only modest changes in magnetic spectral weight associated with the antiferromagnetic ordering at T$_N$ = 32 K. Excitation in higher energy transfer ($\gtrapprox k_{\textrm{B}}T_{\textrm{N}}$) essentially remains the same across the phase transition.\\

We can relate these parameters with measurable quantities via SCR theory \cite{Moriya1987,Moriya1994,Bao1996}. The linear specific heat contribution from spin fluctuation is estimated to be $\gamma \equiv 3 \alpha \pi \textrm{N} \textrm{k}_{B}^{2}/(2\Gamma q_{B}^{2})$, where $\alpha$ = 2 and $q_{B}$ = $(6\pi^{2}/V_{0})^{1/3}$. Here, V$_{0}$ is the volume per a uranium ion and $q_{B}$ is the cut-off wave vector. The Neel temperature, T$_{N}$, can be obtained by $k_{B}\textrm{T}_{N}=0.7137 (\textrm{A}\Gamma^{1/2}\textrm{M}_{Q}^{2}/V_{0})^{2/3}$ where $\textrm{A}\equiv(\chi_{Q}\kappa^{2})^{-1}$ and $\textrm{M}_{Q}$ being staggered magnetic moment.\\

For example, from fit parameters at 50 K with E$_i$ = 100 meV, the T$_{N}$ and linear specific heat coefficient can be calculated to be 2.12 K and 1640 mJ mol$^{-1}$K$^{-2}$ for H-direction and 0.53 K and 7030 mJ mol$^{-1}$K$^{-2}$ for L-direction. These values disagree with measurements, reported in the literature. According to Ref. \cite{Steeman1990}, the linear electronic contribution to the specific heat is 32 mJ mol$^{-1}$K$^{-2}$ or enhanced value of 122 mJ mol$^{-1}$K$^{-2}$ when different phonon model is used. This indicates that while the RPA functional form can reproduce the overall magnetic fluctuation spectrum, the physics in UPt$_{2}$Si$_{2}$ cannot be fully described by the SCR theory. This result is not completely unexpected as SCR theory is assuming weakly itinerant antiferromagnets, where the electronic bandwidth and the Fermi energy scale are much greater than the magnetic energy \cite{Knafo}.\\


\section{VI. Estimation of Bandwidth in the itinerant-electron system}

The RPA expression for magnetic spectral weight in itinerant-electron model is,
\begin{equation}
\chi^{\prime \prime} (q, E) = \frac{\chi_{Q}}{1+(q/\kappa)^2} \frac{\Gamma(q)E}{E^2+\Gamma(q)} ,
\label{eq:IZ1}
\end{equation}
where $\Gamma (q) = \gamma (q^2 + \kappa^2)$. \\ 

The energy-integrated magnetic spectral weight, $\chi^{\prime \prime}(q)$, which (at T=0) gives the static structure factor, or single time correlation function, $S(q)$, diverges unless an energy cut-off, $\Lambda$, is introduced. Then,

\begin{equation}
\begin{split}
S(q) & = \int_{0}^{\Lambda} \chi^{\prime \prime}(q, E)dE = \frac{\chi_{Q}}{1+\left(q/\kappa\right)^2} \int_{0}^{\Lambda} \frac{E/\Gamma(q)}{\left(E/\Gamma(q)\right)^2+1}dE \\
     & = \frac{\chi_{Q}\Gamma(q)}{1+\left(q/\kappa\right)^2} \frac{1}{2} \rm{ln}\left(\left( \frac{\Lambda}{\Gamma(q)}\right)^2+1\right) \approx \frac{\chi_{Q}\Gamma(q)}{1+\left(q/\kappa\right)^2} \rm{ln} \left(\frac{\Lambda}{\Gamma(q)}\right) \\
		 & = \chi_{Q} \gamma_{A} \kappa^2 \left[\rm{ln}\left(\frac{\Lambda}{\gamma_{A}\kappa^2}\right) - \rm{ln}\left(1+\left(q/\kappa\right)^2\right)\right],
\label{eq:IZ2}
\end{split}
\end{equation}

where we have neglected 1 compared to $(\Lambda/\Gamma(q))^2 \gg 1$, which is a reasonable approximation for $\Lambda/\Gamma(q) \gtrapprox 2$. Integrating the above spectral weight in q over the Brilluoin zone yields, 

\begin{equation}
S_{tot}=\int_{-c^*/2}^{+c^*/2} \int_{0}^{\Lambda} \chi^{\prime \prime}(q, E)dE \frac{dq}{c^*} ~\approx \chi_{Q}\gamma_{A}\kappa^2 \left[\rm{ln}\left(\frac{\Lambda}{\gamma_{A}\kappa^2}\right)\right],
\label{eq:IZ3}
\end{equation}

where we have neglected,

\begin{equation}
\begin{split}
\int_{-c^*/2}^{+c^*/2} \rm{ln} \left( 1+\left(q/\kappa\right)^2 \right) \frac{dq}{c^*} & = \frac{\kappa}{c^*} \int_{-c^*/(2\kappa)}^{+c^*/(2\kappa)} \rm{ln} \left(1+x^2\right) dx \\
& = \frac{2\kappa}{c^*} \left[ \frac{c^*}{2\kappa} \left( \rm{ln} \left( 1+\left(\frac{c^*}{2\kappa}\right)^2 \right) -2\right) +2~ \rm{tan^{-1}}\frac{c^*}{2\kappa} \right] \\
& = \rm{ln} \left( 1+\left(\frac{c^*}{2\kappa}\right)^2 \right) -2~ +2~ \frac{2\kappa}{c^*} \rm{tan^{-1}}\frac{c^*}{2\kappa} \\
& = \left( \frac{c^*}{2\kappa} \right)^2 -2 +2 \left(1-\frac{1}{3} \left( \frac{c^*}{2\kappa} \right)^2 \right) + O \left( \left( \frac{c^*}{2\kappa} \right)^4 \right) \\
& = \frac{1}{3} \left( \frac{c^*}{2\kappa} \right)^2 + O \left( \left( \frac{c^*}{2\kappa} \right)^4 \right) \ll 1
\label{eq:IZ4}
\end{split}
\end{equation}

for $\kappa/c^{*} \gtrapprox 1$ (correlation is at least about one lattice unit, which holds for UPt$_2$Si$_2$) compared to the leading logarithm, i. e., assumed

\begin{equation}
\frac{1}{3} \left(\frac{c^*}{2\kappa}\right) \ll \rm{ln} \left(\frac{\Lambda}{\gamma_{A}\kappa^{2}}\right).
\label{eq:IZ5}
\end{equation}

The above expression for the integral magnetic spectral weight leads to the following estimate of the itinerant-electron bandwidth, $\Lambda$, 

\begin{equation}
\Lambda \approx \gamma_{A} \kappa^{2} \rm{exp} \left( \frac{S_{tot}}{\chi_{Q}\gamma_{A}\kappa^{2}} \right)
\label{eq:IZ6}
\end{equation}

Substituting here values from Table 1 and $S_{tot} \approx 15.9 ~\mu_{B}^{2}$ at 5 K and $\approx 13.6 ~\mu_{B}^{2}$ at 50 K, we obtain $\Lambda \approx 1 \sim 2 $ eV. This compares favorably with the energy range where the spin-polarized electronic band structure is observed in the DFT calculations of Elgazzar, et.al., Ref. 9 of the main text. 

\section{V. Bond Energy From First Moment Sum Rule}

\begin{figure}[htp]
\centering
\includegraphics[width=0.7\hsize]{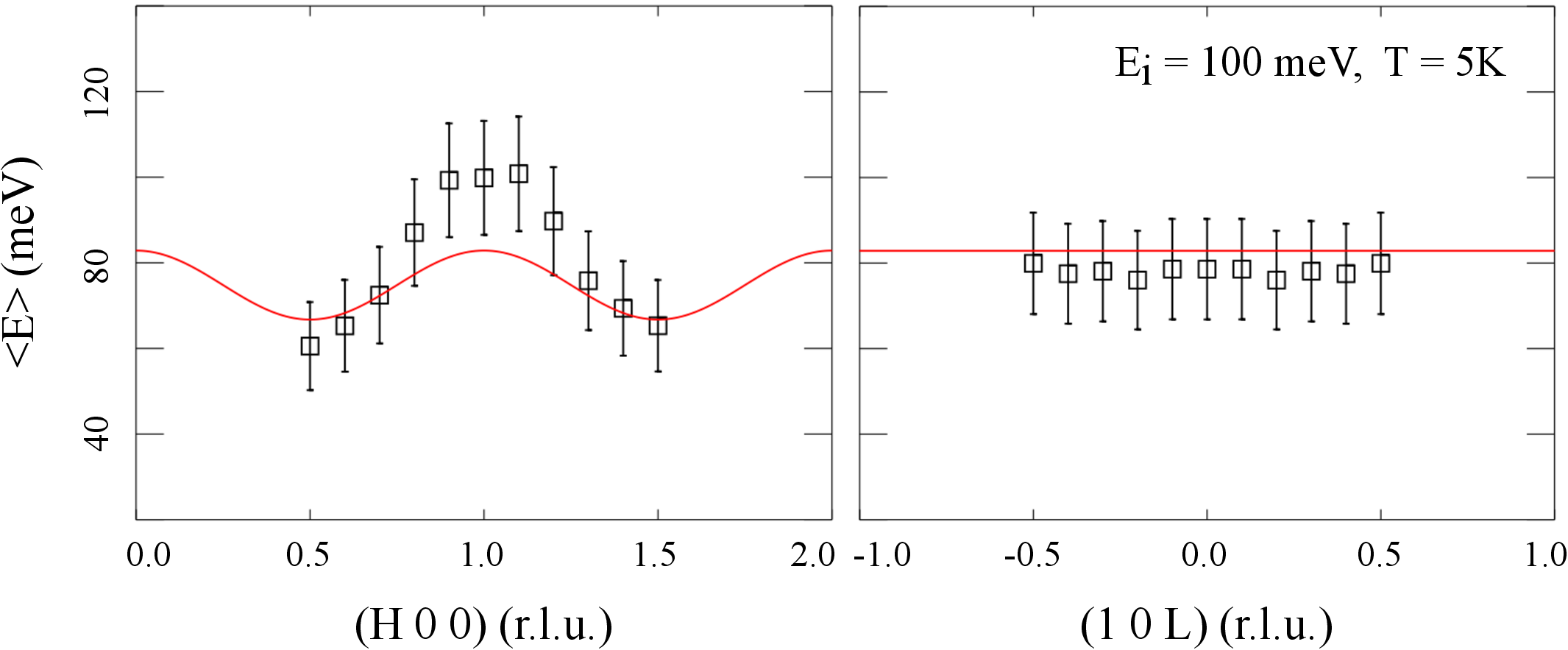}
\caption{Momentum transfer dependence of first moment, $\left\langle \textrm{E} \right\rangle$, along (H 0 0) and (1 0 L) directions. Black square symbol represent the data and red curve shows the fit to the Eq. \ref{eq:sumrule2}.}
\label{fig:S6}
\end{figure}

Information about the bond strength, $\sum_{\textbf{d}} J_{\textbf{d}} \left\langle \textbf{S}_{0} \cdot \textbf{S}_{\textbf{d}} \right\rangle$ can be obtained in a model-independent way using the first moment sum rule of the neutron scattering \cite{MNS, Stone}. Here, $J_{\textbf{d}}$ is the exchange coupling and $\left\langle \textbf{S}_{0} \cdot \textbf{S}_{\textbf{d}} \right\rangle$ is site-independent lattice averaged two-spin correlation energy between magnetic moments connected with lattice vector $\textbf{d}$. For a Heisenberg spin system with a single-ion anisotropy, the expression becomes \cite{MNS} \\

\begin{equation}
\begin{split}
\langle E_{Q} \rangle=& \int^{+\infty}_{-\infty} E ~ S^{\alpha\alpha}(\textbf{Q}, E) dE \\
                     =& -\sum_{d,\beta} (1-\delta_{\alpha\beta})  
                          [J_{d}(1-\rm{cos}(\textbf{Q}\cdot\textbf{R$_{d}$})\langle S^{\beta}_{0}S^{\beta}_{d}\rangle] +\sum_{i,\beta} (1-\delta_{\alpha\beta}) D_{\beta}[S_{i}(S_{i}+1)-\langle(S^{\alpha}_{i})^{2}\rangle-2\langle(S^{\beta}_{i})^{2}\rangle]] .
\end{split}
\label{eq:sumrule1}
\end{equation}

If we consider in-plane exchange interaction and single-ion anisotropy along c-axis, we obtain 

\begin{equation}
\langle E_{Q} \rangle = -\frac{2}{3}\sum_{d,i} [J_{d}(1-\rm{cos}(\textbf{Q}\cdot\textbf{R$_{d}$}) \langle \textbf{S}_{0}\textbf{S}_{d}\rangle + \textrm{D}_{z} \left(\langle(S^{z}_{i})^{2}\rangle -\frac{1}{2} \left(\langle(S^{x}_{i})^{2}\rangle+\langle(S^{y}_{i})^{2}\rangle\right)\right)] = -\frac{2}{3} [E_{a}(1-\rm{cos}(2\pi H)) + E_{D} ]
\label{eq:sumrule2}
\end{equation}

where $\textrm{E}_{a} \equiv \textrm{J}_{a}\langle \textbf{S}_{0}\textbf{S}_{d}\rangle$ and $\textrm{E}_{D} \equiv \textrm{D}_{z} \left[ \langle(S^{z})^{2}\rangle -0.5\left(\langle(S^{x})^{2}\rangle+\langle(S^{y})^{2}\rangle\right) \right]$. This equation relates the first moment of the dynamical structure factor with the expectation value of the exchange bond energy, which is an good estimate of the energy scale of magnetic excitation.\\

We can calculate the momentum transfer dependence of first moment along (H 0 L) direction shown as a red curve in Fig. \ref{fig:S6} from Eq. \ref{eq:sumrule2} and obtain the bond energy per Uranium, $\textrm{E}_{a}$ = 12(3) meV and $\textrm{E}_{D}$ = -124(3) meV. It should be noted that, just like when estimating the total magnetic moment by the zeroth order sum rule, the accessible range around $\rm{Q}_{M}$ =(1 0 0) is kinematically limited and other equivalent high-Q positions suffer from phonon contamination. The first moment, therefore, has been obtained by fitting the spectrum to Damped Harmonic Oscillator (DHO), and the range of interest was limited around $\rm{Q}_{M}$, which is not sufficient to make several sinusoidal oscillations.\\ 

We find that the largest and most important magnetic energy is that of the local magnetic anisotropy resulting from the crystal electric field splitting of the U J-multiplet. While the obtained values should be understood with a grain of salt, the fit suggests that bond energy is positive, that is, the exchange interaction is frustrated. It actually increases the ground state energy. This suggest that the gain in kinetic energy due to the itinerancy beats the loss of the exchange energy: the ferromagnetic in-layer structure is not favored by the effective exchange, which is probably AFM, but by the kinetic energy of the itinerant charges, which favors FM fluctuations. Therefore we find that there is an itinerant electron contribution, unaccounted in the local-moment picture, which lowers the energy of the observed magnetic correlation. 

\section{VI. Wider Range Temperature Dependence at Different Wavevectors}

Here we show temperature dependence of generalized susceptibility at several wavevectors up to 320K, the highest temperature available for the CCR used. Data presented here are from SEQUOIA with E$_{i}$ = 50 meV.\\

\begin{figure}[hbp]
\centering
\includegraphics[width=0.95\hsize]{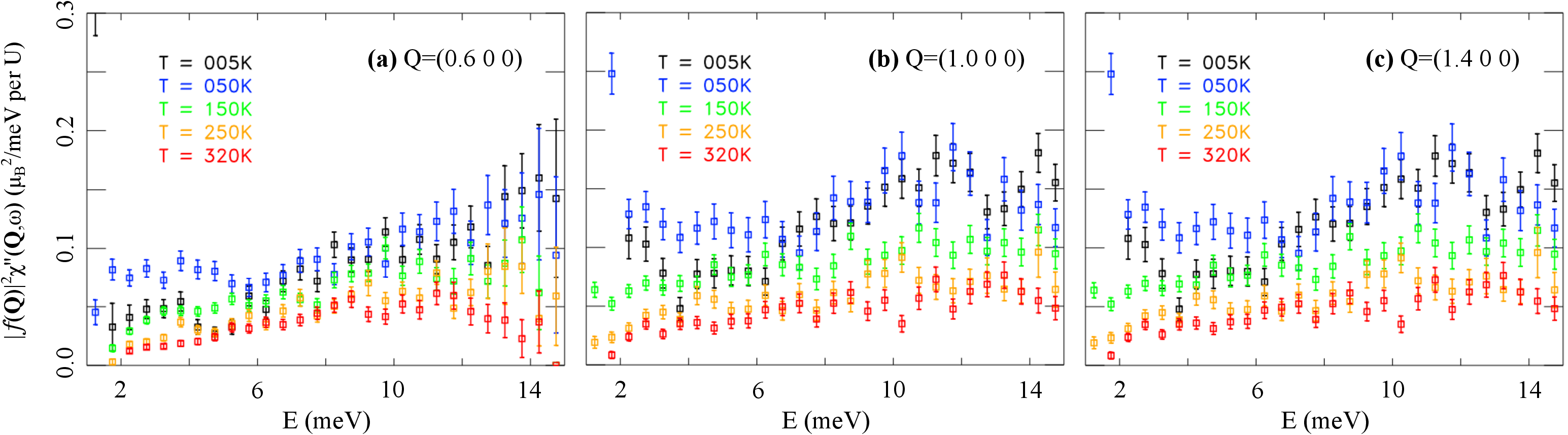}
\caption{Wide range temperature dependence of the generalized susceptibility at wave-vector (a) (0.6 0 0), (b) (1.0 0 0), and (c) (1.4 0 0) obtained from the TOF spectrometer, SEQUOIA with E$_i$ = 50 meV.}
\label{fig:S2}
\end{figure}

Fig. \ref{fig:S2} (a-c) shows temperature dependence for different wavevectors of (H 0 0) with H = 0.6, 1.0, and 1.4, respectively. Here, the integration range are H-range [H-0.1, H+0.1], K-range [-0.2, 0.2], and L-range [-0.3, 0.3]. Their temperature dependence are similar: above the Neel temperature, the low energy gap is filled by spectral re-distribution and the overall magnetic susceptibility keeps decreasing as temperature increases. \\

\begin{figure}[htp]
\centering
\includegraphics[width=0.95\hsize]{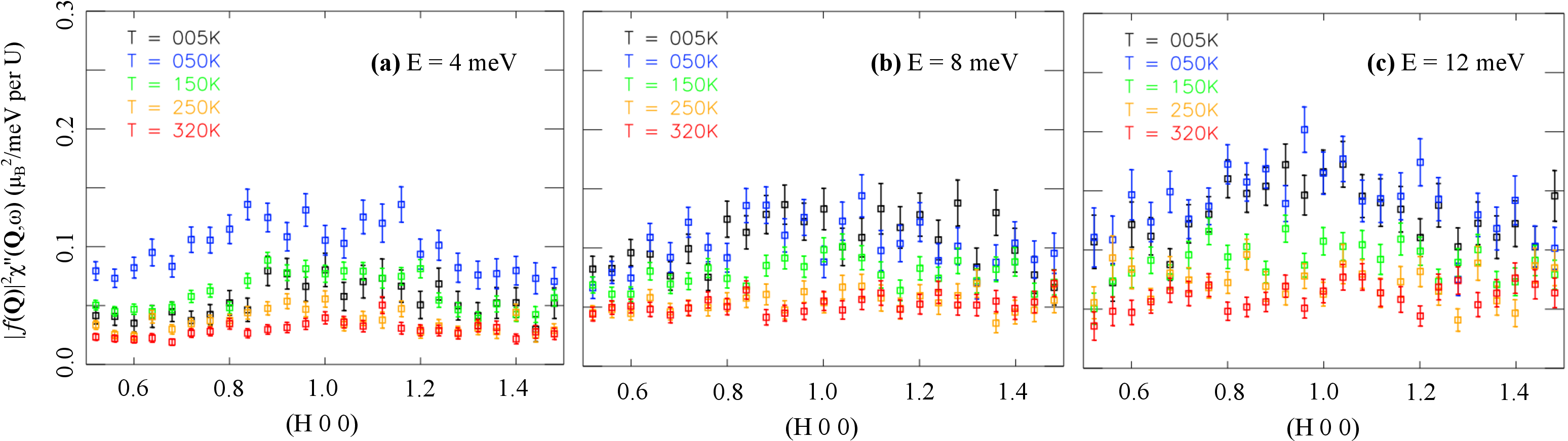}
\caption{Generalized susceptibility along (H 0 0) direction at the energy ransfer of (a) 4 meV, (b) 8 meV, and (c) 12 meV. Data are obtained from the TOF spectrometer, SEQUOIA with E$_i$ = 50 meV.}
\label{fig:S3}
\end{figure}

While the generalized susceptibility is reduced at high temperature, it still survives up to $\approx$ 10 T$_N$, suggesting a strong magnetic correlation. It seems to vanish around 300 K, which can be more clearly visible when we make a constant-energy cut along H-direction as shown in Fig. \ref{fig:S3}. Fig. \ref{fig:S3} (a) represent temperature dependence of susceptibility in the spin gap, i.e., at E = 4 meV. The intensity increase abruptly just above T$_{\textrm{N}}$ and gradually decrease. Fig. \ref{fig:S3} (b, c) show the intensity change of the resonance at E = 8 meV and 12 meV, respectively. Energy integration range here is [E-1, E+1]. The susceptibility reaches background level around the room temperature.

\section{VII.Anisotropy in the magnetic excitation}

\begin{figure}[htp]
\centering
\includegraphics[width=1.00\hsize]{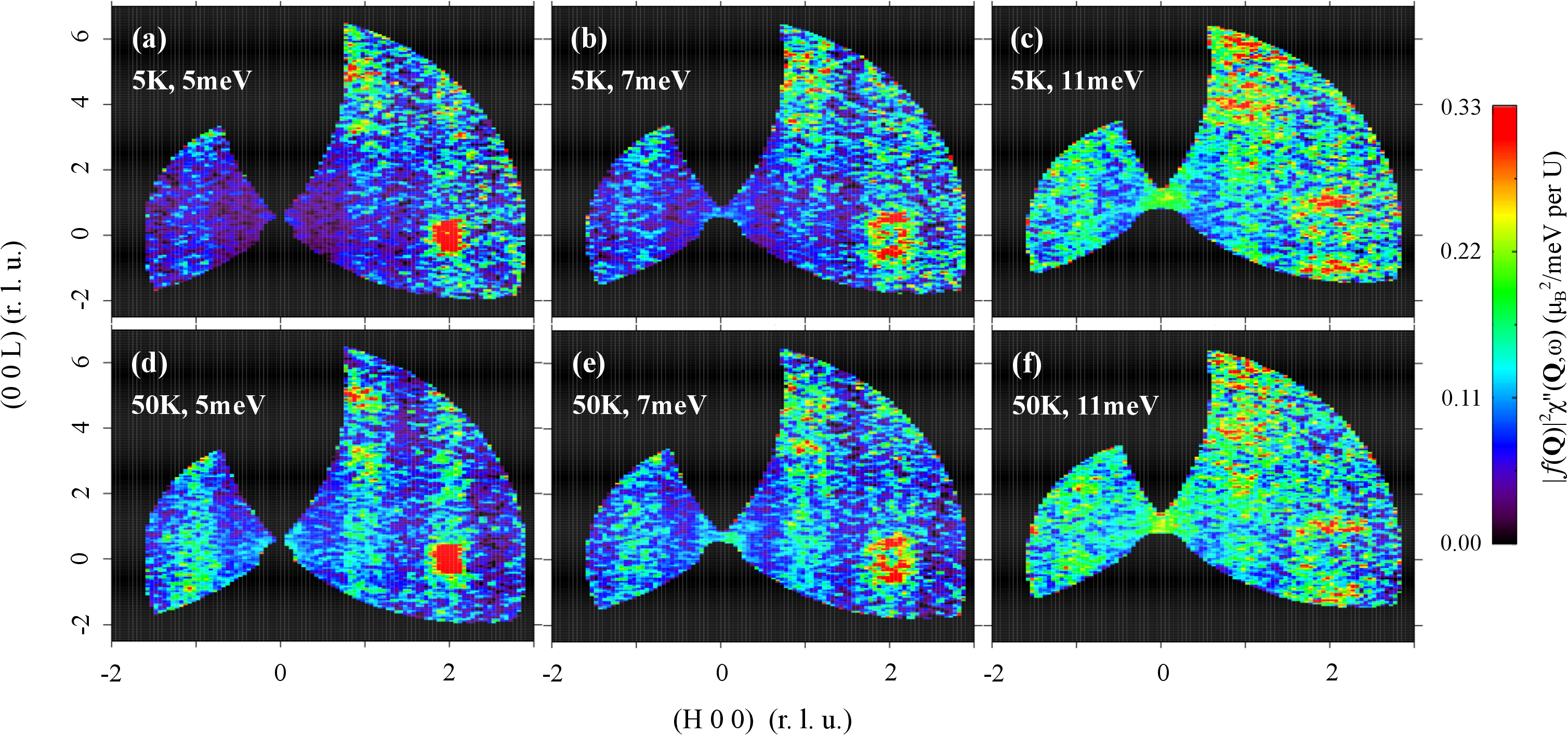}
\caption{Generalized susceptibility at constant energy transfer along H- and L- direction. Data are obtained from the TOF spectrometer, SEQUOIA with E$_i$ = 50 meV.}
\label{fig:S4}
\end{figure}

Anisotropy of magnetic excitation in reciprocal space is best represented when we make a constant-energy slice of Q-E volume. Figure \ref{fig:S4} (a-c) shows map of generalized susceptibility along H- and L- directions at energy transfer of 5, 7, 11 meV at 5 K. Magnetic excitation forms ridge due to dispersionless nature along L-direction. Strong circular excitation around Q = (2 0 0) is from phonon.\\

Above the Neel temperature, 50 K, the magnetic excitation below the spin gap ($\approx$ 7 meV) is enhanced as shown in \ref{fig:S4} (d-f), but anisotropic nature still persists in the correlated paramagnetic excitation.\\

\section{VIII.Absence of spinwave in larger Q-E space}

\begin{figure}[htp]
\centering
\includegraphics[width=0.75\hsize]{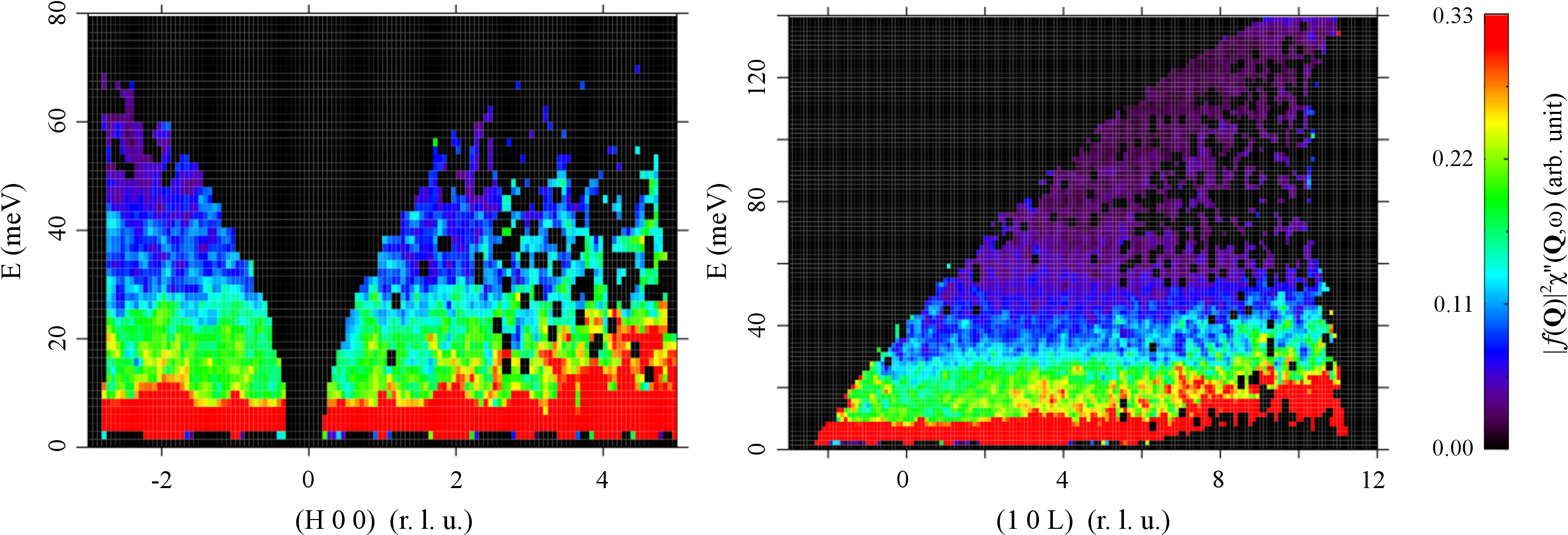}
\caption{Generalized susceptibility along (a) H- and (b) L- direction at 5 K with E$_i$ = 150 meV. Data are obtained from the TOF spectrometer, SEQUOIA, and have not normalized by vanadium standard. }
\label{fig:S5}
\end{figure}

We have explored momentum-energy space with higher incident energy, E$_i$ = 150 meV, to see if there is any spin wave that was not captured. No well-defined coherent spin fluctuation is observed confirming our assertion on the diffusive nature of magnetic excitation.
